\documentclass[preprint,3p,times,twocolumn]{elsarticle}

\usepackage{graphicx}

\usepackage{amssymb}

\usepackage{lineno}

\begin{document}

\begin{frontmatter}

\title{Design study of a photon beamline for a soft X-ray FEL driven by high gradient acceleration at EuPRAXIA@SPARC\_LAB}



\author[lnf]{Fabio Villa \corref{coraut}}
\cortext[coraut]{Corresponding author }
\ead{fabio.villa@lnf.infn.it}
\author[tvuni]{Alessandro Cianchi}
\author[cnr]{Marcello Coreno}
\author[lnf]{Sultan Dabagov}
\author[lnf,cnr,ricmass]{Augusto Marcelli}
\author[tvuni]{Velia Minicozzi}
\author[tvuni]{Silvia Morante}
\author[tvinfn]{Francesco Stellato}

\address[lnf]{Laboratori Nazionali di Frascati - INFN, via E. Fermi 40, 00044 Frascati (RM), Italy}
\address[tvuni]{Dipartimento di Fisica, Università di Roma Tor Vergata \& INFN, Via della Ricerca Scientifica 1, 00133 – Rome - Italy}
\address[cnr]{ISM-CNR, Basovizza Area Science Park, Elettra Lab, 34149 Trieste - Italy}
\address[ricmass]{RICMASS, Rome International Center for Materials Science Superstripes, 00185 Rome, Italy}
\address[tvinfn]{INFN sez. Roma Tor Vergata, Via della Ricerca Scientifica 1, 00133 – Rome - Italy}

\begin{abstract}
We are proposing a facility based on high gradient acceleration via x-band RF structures and plasma acceleration.
We plan to reach an electron energy of the order of 1 GeV, suitable to drive a Free Electron Laser for applications in the so called “water window” (2-4 nm). 
A conceptual design of the beamline, from the photon beam from the undulators to the user experimental chamber, mainly focusing on diagnostic, manipulation and transport of the radiation is presented and discussed. 
We also briefly outline a user end station for coherent imaging, laser ablation and pump-probe experiments. 
\end{abstract}

\begin{keyword}
Photon beamline \sep soft X-ray \sep EuPRAXIA@SPARC\_LAB

\end{keyword}

\end{frontmatter}



\section{Introduction}
\label{intro}

\begin{figure*}[hbt]
\centering
\includegraphics[width=0.95\textwidth]{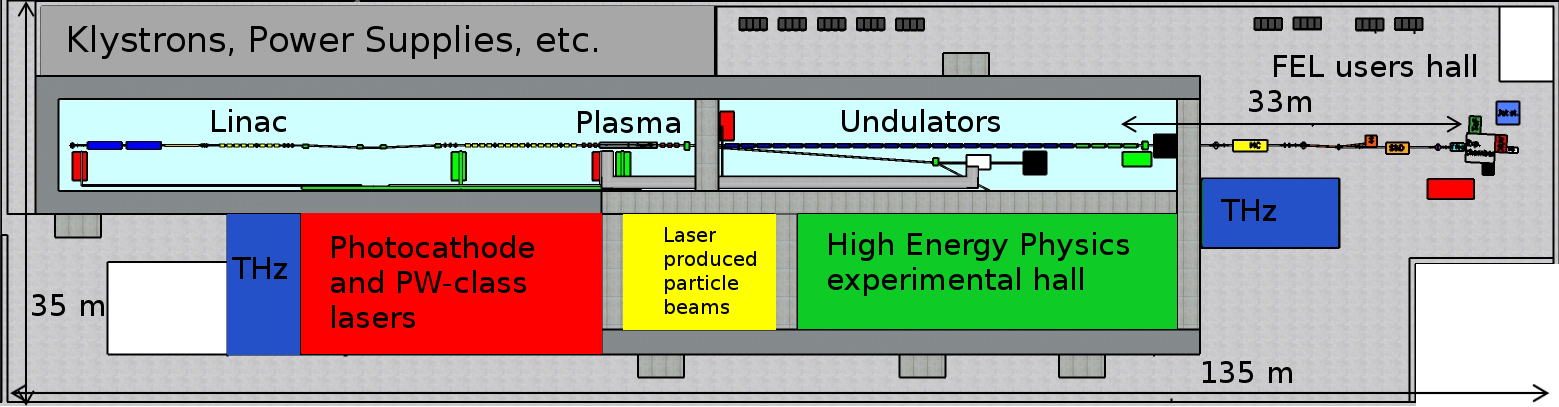}
\caption{Scheme of the EuPRAXIA@SPARC\_LAB facility.}
\label{fig:eusparc}
\end{figure*}

An increasing number of XUV and X-ray facilities based on Free Electron Laser (FEL) are planned.
Key elements of FEL radiation are the high peak brilliance that can be higher than $10^{30}\, photons \, s^{-1} \, mrad^{-2} \, mm^{-2} \, 0.1\%\, bandwidth$ and the short pulse duration, which is of the order of tens of femtoseconds. 
The major disadvantage of X-ray FEL facilities is the large amount of space required for electron acceleration, undulators and photon beamlines, in the order of many hundreds of meters, thus making present FELs available only on national scale and above.  
Plasma Wakefield Acceleration (PWFA) is a promising technique to greatly reduce the required space for beam acceleration, though attaining accelerating gradient stronger than 1 GV/m. 
Up to now, the produced beam quality in term of emittance and energy spread is not yet comparable to conventional RF acceleration methods, while many attempts to close the gap are planned and ongoing. 
We present here the photon beamline of the $EuPRAXIA@SPARC\_LAB$ \cite{eusparc} project.
It consists in an high brightness X-band linac, a stage of PWFA and a FEL. 
This project is intended to be consistent with the EuPRAXIA Design Study requests. 
In figure \ref{fig:eusparc} the possible schematic layout of the EuPRAXIA@SPARC\_LAB facility is presented. 

The experimental activity will be focused on the realization of a plasma driven short wavelength FEL with one user beam line, according to the beam parameters reported in Table \ref{tab:eusparc}. 
The first foreseen FEL operational mode is based on the Self Amplification of Spontaneous Radiation (SASE) mechanism \cite{sase}.
Other schemes, like seeded and higher harmonic generation configurations, will be also investigated. 
In particular we investigated the possibility to fulfill the 1 GeV EuPRAXIA scenario by using plasma acceleration driven by laser or particles, but also the possibility to drive the FEL with higher charge per bunch (100 - 200 pC) in a conventional configuration, exploiting the full X-band RF linac energy (1 GeV) without using the plasma module.

\begin{table*}[htb]
\label{tab:eusparc}
\centering
\begin{tabular}{|l c | l c|}
\hline
\multicolumn{2}{|c|}{\textbf{Electron and undulator parameters}}
 &  \multicolumn{2}{|c|}{\textbf{Radiation parameters}} \\
\hline
Energy & 0.8--1.2 GeV & Wavelength & 2--4 nm  \\
Energy spread & 0.1\% & Bandwidth & 0.15\%  \\
Emittance & 0.5 mm mrad & Dimensions & 0.22 mm  \\
Peak current & 2--3 kA & Divergence & 25 $\mu$rad  \\
Und. period & 15 mm & Photon per pulse & 1--$5\cdot 10^{12}$  \\
K & 1--1.45 & Duration & 5-50 fs rms  \\
\hline
\end{tabular}
\caption{EuPRAXIA@SPARC\_LAB parameters}
\end{table*}

\section{Photon beamline}
\label{pbl}

\begin{figure*}[hbt]
\centering
\includegraphics[width=0.95\textwidth]{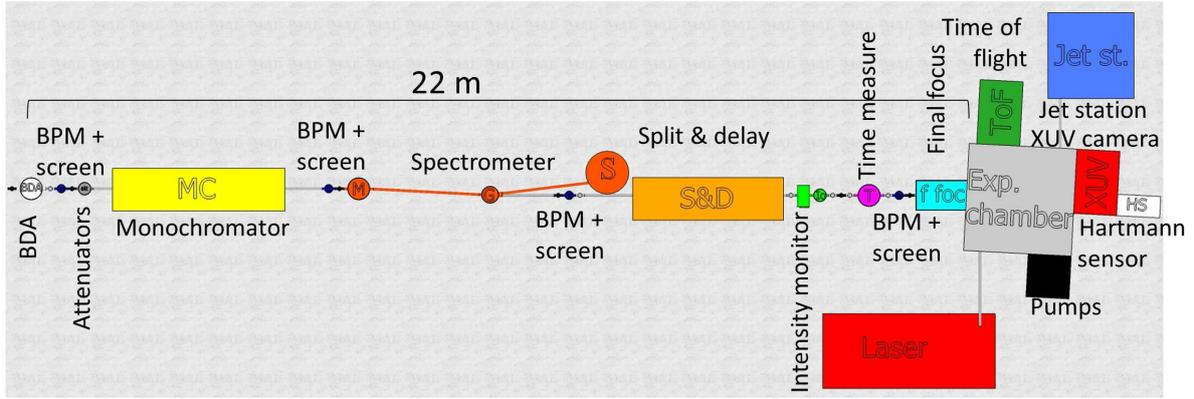}
\caption{Scheme of the photon beamline. The distance of the experimental chamber from the undulators is about 33 m.}
\label{fig:pbl}
\end{figure*}

The SASE radiation presents strong shot-to-shot fluctuations in intensity, spectrum and position. 
The radiation diagnostics should therefore be single-shot and not-intercepting whenever possible. 
The beam will be characterized by measuring its dimensions, coherence and positions both in transverse (section \ref{pbl:trans}) and longitudinal (section \ref{pbl:long}) directions, its spectrum and its intensity (section \ref{pbl:ene}). 
The beamline will also be capable to optimize the beam for the running experiment, to allow the fine tune of some characteristics.
A scheme of the beamline is presented in figure \ref{fig:pbl}

\subsection{Transverse measure and control}
\label{pbl:trans}

The first element on the beamline is the beam defining aperture (BDA). 
Mechanically, the BDA is formed by two trunks of pyramid, where the central aperture of each trunk will be 20x20 $mm^2$. They are independently movable to select the effective aperture up to complete closure.
The aperture will be closed up to having its edges on the tails of the photon beam, so the beam will be almost unperturbed while the following optics and elements will be preserved by large fluctuations or accidental misalignment of the beam position or direction. 
Since the FEL radiation emitted from the undulators contains an intense coherent emission, with an angular divergence of few tens of $\mu$rads, surrounded by a broad spontaneous distribution with a larger angular divergence, this aperture will also act as a collimator.

Along the line several Beam Position Monitors (BPMs) will be installed to monitor the trajectory of the photons.
The BPMs will be based on the interception of the tails of the photon beam transversal intensity distribution by four metallic blades collecting a drain current. 
A polarized plate with a large hole compared to the beam can help to collect the stripped electrons and clean the current signal. 
The expected spatial resolution is determined by the accuracy in measuring currents generated on the blades and by the minimum mechanical step of the motors controlling the travel of the blades. 
In FERMI BPMs the current accuracy is about $10^{-6}$ A (AH401 picoammeter \cite{zangrando}) while on the step the accuracy is about 1 $\mu m$. 
To avoid ablation due to the high peak energy, the blades will be tilted by about $20^\circ$. 
Each blade may travel few cm, and a complete closure in both directions is possible.  
By reading simultaneously the four currents, it is possible to determine pulse-by-pulse the relative displacement of each pulse with a spatial resolution of about 2 $\mu m$ rms. 

We will measure the transverse dimension of the photon beam via a scintillating screen. 
The material commonly used as scintillator is YAG:Ce.
It is cost-effective and it has a high yield of light (about 8 photons/keV \cite{blm-xfel}). 
However, the photon flux at high intensity may damage the crystal, and other materials with a higher damage threshold, such as pc-CVD diamond \cite{blm-xfel}, can be envisaged. 
Behind the screen, a $45^\circ$ mirror and a camera outside the screen vacuum chamber in a setup similar to those used for electron diagnostics will be available.

\begin{figure}[hbt]
\centering
\includegraphics[width=0.475\textwidth]{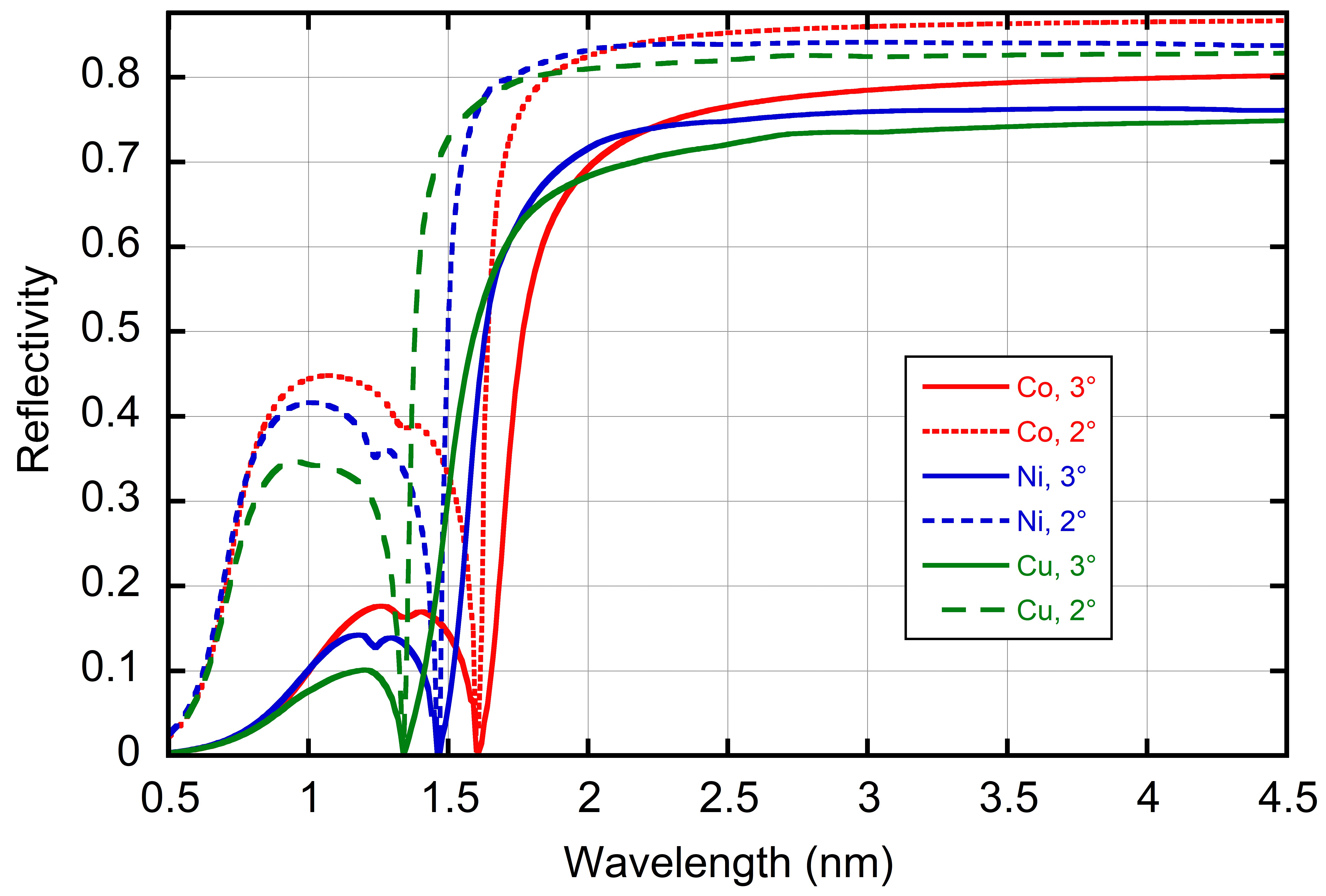}
\caption{Reflectivity of thick layers of different metals (Cu, Ni, Co) at 2$^\circ$ and 3$^\circ$ grazing angle, extrapolated using data from \cite{reflectivity}. The reflectivity remains almost constant at longer wavelengths}
\label{fig:reflectivity}

\end{figure}

Few mirrors are required in the beamline: for steering the beam away from the undulator line, for a split and delay system, for the monochromator and for the final beam focusing.
Each mirror will have two angular degrees of freedom and an insertion control. 
The mirror will be inclined to an angle of $90^\circ - \alpha$ (where $\alpha$ is the grazing angle of the mirror, in our case between 1 - 3$^\circ$) in the horizontal plane with respect to the incoming beam direction. 
Their substrate will be made by fused silica, while the coating material is depending on the wavelength used \cite{coatings}. 
We are studying different coatings that have a reflectivity $>$ 70\% at 3$^\circ$ grazing angle for almost all the spectral range 2-20 nm also when used in a single layer configuration to maximize durability and minimize the costs (Figure \ref{fig:reflectivity}).

\begin{figure}[hbt]
\centering
\includegraphics[width=0.425\textwidth]{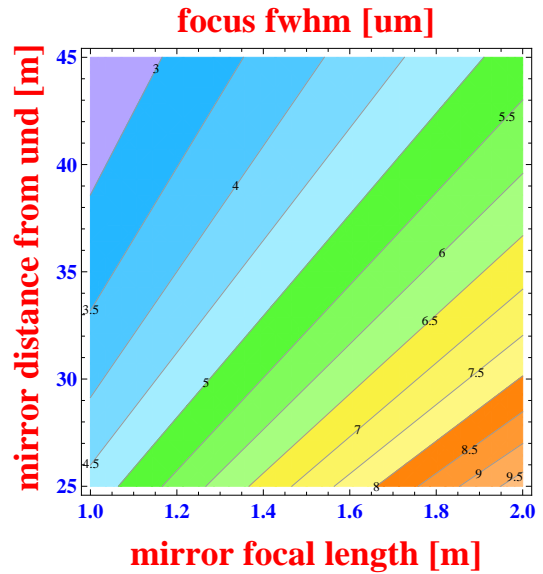}
\caption{Preliminary focal spot dimensions, calculated with ideal Gaussian beam and aberration free focusing, as a function of mirror focal length and distance from the undulators. The black dot represents the actual working point, 33 m from the undulator and with 1.5 m focal length, of about 5.5 $\mu$m FWHM.}
\label{fig:focus}
\end{figure}

For what concerns the focusing device, we plan to use two mirrors (spherical or plane elliptical) in a Kirkpatrick-Baez configuration. 
The curvature can be manufactured or implemented directly by slightly bending the mirrors \cite{KB-fermi}. 
The equivalent focal length can be in the range of few meters or less (as a reference, the bent mirror at FERMI can have a minimal focal length of $\sim$1.2 m and a minimum spot size of 2x3 $\mu m^2$ \cite{photbeamline-allaria}). 
Depending on radiation parameters, distance from the undulators and mirror focus, the final spot size will be in the order of some microns rms (Figure \ref{fig:focus}).
Moreover, multi channel plates (MCPs) can be considered as they offer promising results for coherent X-ray imaging. 
The MCP parameters are very flexible thus becoming attractive to design novel X-ray optics \cite{Dabagov, Mazuritskiy}.

Measurement of transverse coherence measurement will be possible, exploiting the visibility of an interference pattern from a double slit \cite{doubleslit} or, as demonstrated also at SPARC\_LAB, using the speckle pattern from small spheres \cite{alaimo1, alaimo2}. 
We will introduce the slits (or the spheres) in a dedicated chamber after the spectrometer, depicted in section \ref{pbl:ene}, and we will use a screen near to the experimental chamber to visualize the pattern.

\subsection{Longitudinal measure and control}
\label{pbl:long}

Commercially available streak cameras have resolution in the order of few hundred fs. 
While the temporal length can be reconstructed by the electron longitudinal phase space after the undulators, as in \cite{Behrens}, although with possible systematic errors, a dedicated diagnostic tailored to our photon characteristics could be investigated. 
Many technologies have been considered for the characterization of short X-ray pulses \cite{Dusterer}, based, for example, on interferometry \cite{Dininno, Kayser}, transient reflectivity \cite{Riedel}, cross-correlation \cite{xcorr-fermi} or THz streaking \cite{thzstreak}. 
Interferometry requires a dedicated multishot measurement, so the SASE shot-to-shot fluctuation will be averaged and only averaged parameters will be accessible. 
Transient reflectivity and crosscorrelation can be single-shot measurements, but they intercept the beam. 
THz streaking of gas ionization has the potential to be a single-shot non intercepting diagnostic, but is still under development.  

The FEL wavefront will be measured by means of a Hartmann sensor, that is composed by a mask with several holes and a scintillating screen. 
The displacement between the image position of the holes respect to a reference plane wave will give information on the wavefront angle. 
Such instrument is routinely implemented at FEL facilities, such as FERMI \cite{Svetina} and FLASH. 

Commercially available fast photodiodes have temporal resolution of few tens of ps, to increase the time resolution and to allow monitoring the time jitter and arrival time with tens fs resolution other techniques are required. 
We plan to develop the THz streaking measurement to obtain also the arrival time with high resolution \cite{thzstreak}, comparing the electron time of flight with the THz pulse. 
Alternative methods can be considered, which are based on spectral encoding \cite{Bionta} or transient reflectivity \cite{Riedel} using an external reference laser, but both are intercepting single-shot measurements.

As some experiments require very short pulse (i.e. of the order of 10 fs), the pulse length exiting from the undulator should be preserved. 
The chromatic dispersion of the material is very low, as the beam goes through only thin attenuators and low pressure gases. 
The monochromator diffracting gratings, conversely, can induce a large pulse lengthening due to pulse front tilting. 
The path difference between tail and head of the pulse dispersed by one grating is equal to $N$ $m$$\lambda$, where N is the number of grooves illuminated by the beam, $m$ the diffraction order and $\lambda$ the pulse central wavelength. 
For example, the pulse front tilt induced on the diffracted beam by a grating configuration similar to that considered for the spectrometer described in section \ref{pbl:ene} is about 690 fs for a 1.5 mm full width beam. 
If short pulses with high spectral purity are required, it is possible to compensate the pulse front tilt of the monochromator via a pair of gratings working with opposite angles respect to the induced dispersion \cite{Frassetto}, but at the cost of higher losses due to the diffraction efficiency (that is in the order of few tens \% even with high performance gratings \cite{Mcentaffer}). 

We also propose to include a split and delay system in the beamline for XUV pump-probe experiments. 
This system uses the edge of a mirror to split the beam in two components that are then reflected on a delay line and recombined on a final mirror, with a tunable delay typically ranging from 0 to few hundreds fs. 
Such devices are installed both at BL2 at FLASH \cite{Wostmann} and at LCLS \cite{Castagna}.

We plan to measure the transverse coherence length of the beam using an interferometer in Michelson \cite{Hilbert} or Fizeau configuration. 
The decay of the fringe visibility as a function of the arm length and the respective time delay will be used to determine the temporal coherence length. 
This is a multishot intercepting measurement and only the average behavior will be obtained. 
The interferometer can be set inside or just after the final experimental chamber.  

\subsection{Energy measure and control}
\label{pbl:ene}

We will measure the number of photons per pulse using gas-based intensity monitors similar to those used for FLASH \cite{Richter} and FERMI \cite{zangrando}.
The working principle of the intensity monitor is the atomic photo-ionization of a rare gas at low particle density, in the range of $10^{11}$  cm$^{-3}$  (p $\sim 10^{-5}$  mbar). 
The photon beam traveling through a rare gas-filled chamber generates ions and electrons, which are extracted and collected separately. 
This monitor is almost completely transparent due to the low pressure used for the rare gas in the vacuum chamber. 
It has a wide dynamical range and it does not suffer from saturation effects. 
Moreover, it is independent from the beam position fluctuations so it can be used continuously for on-line shot-to-shot intensity measurements. 
This monitor can be calibrated with different sources by using cross-calibrated photodiodes, with an expected precision of 3\% in most of the range (slightly worse at lower energy). 
One of the limiting factors is the ability to read very low current, because the ionizing cross section of commonly used gas, nitrogen, decreases in water window spectral region and the emitted photocurrent is typically low. 
This may require the use of other gas (such as xenon \cite{Richter} or oxygen \cite{Turner}) or other methods of energy detection, e.g. using the spectrometer to obtain a relative intensity measurement.

In order to control the beam intensity, attenuators will be installed in the beamline. 
While gas attenuators have a continuous set of attenuation parameters and generally have a large spectral range, we have to manage photons in a relatively short bandwidth, so thin films can be used instead. 

We will measure the pulse spectrum with a spectrometer based on diffraction gratings \cite{zangrando, Poletto}. 
We plan to use a mirror at near grazing incidence (3$^\circ$) to shift the beam from the undulator line. 
This will be useful to separate the desired radiation pulse from all the other radiations or particles that are propagating on axis.
After the mirror a grating in $0^{th}$ order is going to reflect the beam toward the experimental chamber, whereas its first order can diffract a small portion of the FEL pulse toward a CCD camera.  
The grating will have a groove density between 1200-3600 grooves/mm. 
The separation space between the elements will be determined to avoid radiation propagating through the pipe line, while a first estimation is in the order of few meters. 
After the grating chamber, the CCD camera will be about 2-3 m far from the grating, with an angle in the range of 2 - 7$^\circ$ relative to the reflected beam, depending on groove density and radiation wavelength. 
For a 2400 grooves/mm grating and 3 m of propagation, one camera pixel (about 13 $\mu$m for a back illuminated soft X-ray camera) will cover 2.4 $10^{-4}$ nm (about 1/12 of the expected spectral width) at an angle of about 4.5$^\circ$ from the reflected beam. 
The high quantum efficiency (in the order of 30\%) and the signal-to-noise ratio of the camera require only a small fraction of the photon beam to be diffracted.
The grating will therefore be designed to have a low diffraction efficiency.  

If high spectral purity or very narrow bandwidths are required, we should additionally filter the spectrum. 
For the considered materials and incidence angles, the mirror reflectivity drops sharply below 2 nm, with a smaller reflectivity peak at about 1 nm as reported in Figure \ref{fig:reflectivity}. 
If a more precise or tunable spectrum should be used, we can consider the use of a monochromator \cite{Frassetto} as discussed in section \ref{pbl:long}.

\section{Experimental endstation}
\label{users}

A FEL in the water window with short, coherent and intense pulses is very appealing for different studies, as detailed in \cite{notaextrim}, e.g. 2D and 3D coherent imaging of biological and non-biological samples, nanocluster studies, laser ablation/desorption diagnostics and pump-probe for Raman or four wave mixing spectroscopies.

A dedicated experimental setup will be installed to perform the widest possible class of experiments \cite{notaextrim}.
The experimental chamber will have the possibility to host solid samples on motorized stages and will have the possibility to be connected to a sample delivery apparatus capable of handling also liquid and gaseous targets, such as the one described in \cite{Deponte}.
Systems of this kind have already been successfully used at existing FEL sources FLASH, LCLS and SACLA. 
Different detectors will be located inside the chamber.
A reference example is the CAMP instrument successfully installed at FLASH and LCLS \cite{Struder} whose dimensions are about 2.5x1.5x1.5 m$^3$.
A time of flight spectrometer connected to the experimental chamber will be used to analyze the molecules produced by the sample-beam interaction \cite{Sorokin}.
High power synchronized optical lasers will be considered to allow performing laser pump-FEL probe experiments. 
Dedicated servers and computers will be installed to control the beamline and to allow data storage and processing.
A supporting laboratory, in particular for biological/chemical preparations and manipulations, located next to the beamline will be made available to allow last-minute sample preparations and characterizations.

\section{Conclusions}
\label{conc}

The length of the beamline upstream of the experimental apparatus required to host the above described diagnostics may range from 15 to 25 m, with a distance from the undulators not smaller than 10 - 15 m, mainly depending on photon intensity and optics damage thresholds. 
The expected total transmission of the beamline, when no intercepting devices are operational, is $\sim 35\%$, eventually reduced to $\sim 13\%$ when split and delay line is used. The monochromator will additionally reduce the transmission about two orders of magnitude. Final focus will further reduce the transmission to about 20\% (8\% with split and delay). 
The presented beamline will be capable of fully characterize and manipulate the beam parameters to the experimental chamber, using a compact and efficient line in order to perform high class experiments as described in \cite{notaextrim}. 

This work was supported by the European Union's Horizon 2020 research and innovation programme under grant agreement No. 653782.





\bibliographystyle{elsarticle-num}
\bibliography{bib_proc_eaac17}







\end{document}